\documentclass[amsfonts, 12pt]{article}
\usepackage{amssymb}
\usepackage{graphicx}
\usepackage{url,hyperref}

\begin{document}

\begin{titlepage}
\begin{center}
\vspace{-5in}
\hfill
arXiv:0905.0540 [hep-th]

\vspace{1cm}
{\Large {\bf  Inner Brane: A D3-brane in Nappi-Witten space from an inner group automorphism}}
\vspace{1.5cm}

Yeuk-Kwan E. Cheung\\
\vspace*{3mm}
{\it{
Department of 	Physics, Nanjing University\\
22 Hankou Road, Nanjing 210093, China}}

\vspace*{5mm}
Laurent Freidel\\
\vspace*{3mm}
{\it{Perimeter Institute for Theoretical Physics\\
31 Caroline St North, \\
 Waterloo, ON N2J 2Y5, Canada\\}}

\end{center}

\begin{abstract}
WZW models are  abstract conformal field theories with an infinite dimensional symmetry which accounts for their integrability,  and at the same time they  have a sigma model description of  closed string propagation on group manifolds which, in turn, endows the models with an intuitive geometric meaning.   
We exploit this dual algebraic and geometric property of WZW models to  construct  an explicit example of a   field-dependent   reflection  matrix  for open-strings in  Nappi-Witten model. 
  Demanding the momentum outflow at the boundary to be zero determines a certain combination of the  left and right chiral currents at the boundary.  This same reflection matrix is obtained algebraically  from an inner automorphism,  giving rise to a space-filling D-brane.   Half of the infinite dimensional affine  Kac-Moody  symmetry present in the  closed-string theory is preserved by this  unique combination of the  left and the right chiral currents.  The OPEs of these boundary currents are computed explicitly and they are  showed to obey the same current algebra  as those of  the closed-string chiral currents.  
Different choices of the inner automorphisms correspond to different background gauge field configurations.  
Only those B-field configurations, and the corresponding D-branes, that preserve the diagonal part of the infinite dimensional chiral  algebras are allowed.   In this way the existence of the D-branes in curved spaces is further constrained by the underlying symmetry of the ambient spacetime.

\end{abstract}
\vspace{0.5cm}

KEYWORDS:  D-branes, Current Algebra, WZW model, String Sigma Model

\end{titlepage}

\newcommand{\be}{\begin{equation}}
\newcommand{\ee}{\end{equation}}
\newcommand{\beq}{\begin{equation}}
\newcommand{\eeq}{\end{equation}}
\newcommand{\bea}{\begin{eqnarray}}
\newcommand{\eea}{\end{eqnarray}}
\newcommand{\bary}{\begin{array}}
\newcommand{\eary}{\end{array}}
\newcommand{\nn} {\nonumber \\}
\newcommand{\spa}{\hspace{1cm}}

\newcommand {\eqr} [1]  {{(eq.\,\ref{eq:#1})}}
\newcommand {\Eqr} [1]  {{(Eq.\,\ref{eq:#1})}}
\newcommand{\cB}    {{\mathcal B}}
\newcommand{\cBbar} {{\bar{\cB}}}
\newcommand{\cbar}  {{\bar c}}
\newcommand{\cC}    {{\mathcal{C}}}
\newcommand{\cF}    {{\mathcal F}}
\newcommand{\cH}    {{\mathcal{H}}}
\newcommand{\cg}    {{\mathcal g}}
\newcommand{\cG}    {{\mathcal G}}
\newcommand{\cI}    {{\mathcal I}}
\newcommand{\cJ}    {{\mathcal J}}
\newcommand{\bJ}	{{\bar{J}}}
\newcommand{\cL}    {{\mathcal{L}}}
\newcommand{\cLbar} {{\bar{\cL}}}
\newcommand{\cM}    {{\mathcal M}}
\newcommand{\cMbar} {{\bar{\cM}}}
\newcommand{\cN}    {{\mathcal N}}
\newcommand{\Nbar}  {{\bar N}}
\newcommand{\cO}    {{\mathcal{O}}}
\newcommand{\cP}    {{\mathcal{P}}}
\newcommand{\qbar}  {{\bar q}}
\newcommand{\cR}    {{\mathcal{R}}}
\newcommand{\cS}    {{\mathcal S}}
\newcommand{\cSbar} {{\bar{\cS}}}
\newcommand{\cT}    {{\mathcal T}}
\newcommand{\tbar}  {{\bar t}}

\newcommand{\hV}  {{\hat V}}
\newcommand{\tilV}  {{\widetilde V}}
\newcommand{\vbar}  {{\bar v}}
\newcommand{\vtil}  {{\tilde v}}

\newcommand{\cX}    {{\mathcal X}}
\newcommand{\Wbar}  {{\overline W}}
\newcommand{\bw}    	{\bar{w}}
\newcommand{\cZ}    {{\mathcal Z}}

\newcommand{\zbar}  {{\bar z}}
\newcommand{\Zbar}  {{\overline Z}}
\newcommand{\bz}{\bar{z}}
\newcommand{\ba}{\bar{a}}

\def\tr#1{{{\mathrm tr}\{ {#1} \} }}            
\def\Tr {{\mathrm Tr}}

\newcommand{\dirac} {\not \!}
\newcommand{\diracD} {\not \! \! D}

\newcommand{\pd}    {\partial}
\newcommand{\bpd}      {{\bar \partial}}


\addcontentsline{toc}{section}{Introduction}

Since their discovery~\cite{Dbranes} studying D-branes has been a very important aspect of research in string theory.   While D-branes have simple interpretation as being the hypersurface confining the endpoints of the  open strings from the sigma model point of view, it is much harder to study them in an abstract conformal field theory that does not necessary have any geometric interpretation  for  the  fields as coordinates of a target spacetime.  Important progress has been made throughout the years~\cite{
Polchinski:1987tu,  Abouelsaood:1986gd,  Ooguri:1996ck,  Fuchs:1997fu, Maldacena:2001ky, Giveon:2001uq, Lee:2001gh,  Ponsot} 
especially in the case of  D-branes in WZW models~\cite{
Klimcik:1996hp,  Stanciu:1998ak,  Alekseev:1998mc,  Stanciu:1999nx, Alekseev:1999bs, Felder:1999ka,   
Stanciu:1999id,   Figueroa-O'Farrill:1999ie,  Malda-Oog, Pawelczyk:2000ah, Ishikawa:2000ez,  Bachas:2000fr,  Stanciu:2001vw, Stanciu:2001ug,
Sarkissian:2002ie, Lindstrom:2002mc,  Lindstrom:2002vp, Stanciu:2003bn,  Albertsson:2003va,  Sarkissian:2003yw, 
Giribet:2007wp, Baron:2008qf}.

WZW models~\cite{Witten:WZW}  are very special.   Not only do they have a sigma model description in which the fields take values in a group manifold  of a particular Lie algebra, they are also  conformal field theories  with an infinite dimensional affine Kac-Moody symmetry, where the  current algebra is built upon the very same Lie algebra.    In this way they nicely relate  the algebraic bootstrap approach of the  conformal field theory to the more intuitive geometric approach associated with  the sigma model analysis.  
While closed string propagation on these group manifolds is much studied, much less is known about the properties of open strings on these manifolds.  The   known class of  D-branes were obtained by gluing the left-moving and  the right-moving worldsheet currents in the closed string theory with  a  field independent
``reflection'' matrix, $\cR$,   {\em i.e.} 
${\cJ}_L = \cR {\cJ}_R$, with a notable exception of~\cite{Stanciu:1998ak}.  We would like to  propose using inner automorphisms  as general gluing condition,  giving rise to field-dependent gluing matrices, $\cR$.\footnote{We speak of inner automorphisms of adjoint action by an arbitrary group element, $g$.  In the case of identity,  $e$,  the gluing condition is again  $\cJ_L + \cJ_R=0$,  coincided with the previously studied case and corresponding to the Neumann boundary condition.}
Different choices of the inner automorphisms are related to  different configurations of the background gauge fields coupled to the endpoints of the open strings.   However choosing a consistent background gauge field for the open-strings is far from obvious within the conformal field theoretical  framework.  The intuition   gained  from the  geometric approach serves  well in this regard.   Sigma model analysis of the boundary conditions also straightforwardly  endows  the D-brane with a geometric interpretation.  The underlying  symmetry of the D-branes is, on the other hand, most easily seen by studying  the symmetry preserved by the boundary conformal theory.  The algebraic and geometric approaches  compliment each other and provide a  complete picture of the D-branes.

 So far the last aspect of D-branes, namely what symmetry of the underlying curved spacetime is perserved by the D-branes, has been mostly overlooked.   We would like to advocate the view that the existence of the D-branes are further constrained by underlying symmetry of the ambient (curved) spacetime.  Only those B-field configurations, and the corresponding D-branes, that preserve a subalgebra  of the infinite dimensional chiral  algebras at the boundary are allowed.   Without a good  understanding  of how  the underlying infinite dimensional symmetry  constrains the existence of  the D-branes, the study of D-branes is curved space is not complete.

We exploit this unique  gift  of WZW theories to uncover a new type of field dependent gluing conditions,  illustrating our techniques using  the  Nappi-Witten WZW model~\cite{Nappi:1993ie}, a pp-wave  background supported by a  null NS three-form flux. 
We  show, in this note,  that it is in fact  easy  to generalize to  field-dependent gluing matrices using  the geometric data of  the boundary conditions.   A very special choice of the inner automorphism, adjoint action by the group element,  $e^{uJ}$,  gives rise to a space-filling brane in NW model.  This  gluing condition was first proposed in~\cite{Stanciu:1998ak}.  Stanciu~\cite{Stanciu:1998ak, Stanciu:1999nx, Stanciu:1999id, Figueroa-O'Farrill:1999ie, Stanciu:2001ug} was also the first to stress the importance for the D-branes to preserve part of the underlying affine symmetry.  The construction of the D3-brane whose worldvolume  spanning all four directions of the Nappi-Witten space is as follows.  In Section~1 we construct the boundary conditions from momentum consideration and show that all four directions are Neumann.  Section~2    the reflection matrix relates the left-moving currents with the right-moving ones is obtained through an algebraic method.  The momenta obtained in Section~1 are re-expressed in terms of chiral currents.   This   unique combination of the  left and the right chiral currents  is shown,  in Section 3,  to satisfy the same current algebra as the chiral currents of the closed strings.  The boundary currents hence  preserve  half of the infinite dimensional affine  Kac-Moody  symmetry present in the  closed-string theory.  
The readers can see for themselves that it is  much more straightforward to study D-branes 
if one fully  exploits the dual  geometric and algebraic properties of WZW models. 

Nappi-Witten model receives a lot of attention lately because it is one of the solvable string models in the plane-polarized gravitational waves~\cite{%
Blau:2001ne, Metsaev:2001bj,  Berenstein:2002jq, Metsaev:2002re}.  
Following this realization  D-branes in  these backgrounds have been extensively studied (See~\cite{ppbranes} for a sample of literature.).
Nappi-Witten model, describing closed string propagation in a pp-wave background supported by   light-like NS fluxes, has the added merit of being a WZW model which eventually led to its complete and covariant solution, via a Wakimoto ``free-field''  realization~\cite{Cheung:2003ym},  in which  string vertex operators were constructed and scattering amplitudes for  an arbitrary number of tachyons were  presented.  
(See also~\cite{D'Appollonio:2003dr} for a different way of deriving the three and four point  amplitudes by taking a pp-wave contraction of the $SU(2) \times U(1)$ amplitudes.)
In this note we  will use a space filling D3-brane  in NW model to illustrate our techniques.
 Emphasis is placed on the interplay of the geometric and algebraic approaches.   
The fact  that the free fields obey Neumann boundary condition at the boundaries is made  apparent (in Section 3)
using the covariant  free field realization  introduced in~\cite{Cheung:2003ym}.


\section{Sigma Model Analysis}

We shall begin with the  sigma model side of the story.  Starting with a generic string sigma model (in conformal gauge):
\be
\cL = \int  G_{\mu\nu}\,  \pd X^\mu \cdot \pd X^\nu 
               +\epsilon^{ab} \cF_{\mu\nu} \pd_a X^\mu \pd_b X^\nu
\ee
Upon variation of the sigma model action there are contributions from the boundary terms.   Consistency of the open-string background requires that  such terms vanish.  In the presence of   the B-field, $B_{\mu\nu}$,  and a background gauge field, $F_{\mu\nu}$,  coupling to the end-points of the  open strings   the boundary terms  read
\be
\delta X^\mu \left[ G_{\mu\nu} (X) \, \partial_{\sigma} X^{\nu} +  
\cF_{\mu\nu} (X) \, \partial_{\tau} X^\nu \right]  \big|_\sigma  = 0
\ee
in which only the gauge invariant combination, $\cF = B+ F$, appears.
The boundaries of the worldsheet  are set at  $\sigma=0$ and $\sigma=\pi$.
Writing the above using the  left-moving and right-moving worldsheet derivatives,
  $\pd_\pm = \frac{1}{\sqrt{2}} (\pd_\tau \pm \pd_\sigma)$, we have 
\be	\label{eq: sigmaBC}
\pd_+ X^\mu = \left[ (G + \cF)^{-1} ( G- \cF )\right]_{\mu\nu} \pd_- X^\nu~.
\ee
provided that  $(G+ \cF )$  is invertible, which it will be for  the case at hand.

Neumann condition is equivalent to the requirement that no worldsheet momentum flows out of the boundary of the worldsheet.  The worldsheet momentum  flowing across  a small element,  $d\vec{l}=(d\tau, \, d\sigma)$,  of the boundary of the worldsheet  is given by
\be 	\label{eq: dP}
dP^\mu  =   P^\mu_\tau \, d\sigma -  P^\mu_\sigma \, d\tau
\ee
has to vanish in order for the endpoints of the string to be free.
The components of the worldsheet momentum are defined canonically by
\be
\displaystyle{P^\tau_\mu=\frac{\delta \cL}{\delta \pd_\tau X^\mu}}
                                      \spa \spa
\displaystyle{ P^\sigma_\mu=\frac{\delta \cL}{\delta \pd_\sigma X^\mu}}
 \ee
 To read off  the gluing condition for the Neumann directions,  say  in the ``open-string picture''
{\em i.e.} an open segment terminating at $\sigma=0$ and $\sigma=\pi$ at the 
D-brane~($d\sigma =0$),  we demand no momentum outflow at all values of    $\tau$.  
We therefore must have  
\be
 P^\sigma_\mu \, = \,G_{\mu\nu} (X) \, \partial_{\sigma} X^{\nu} +  
\cF_{\mu\nu} (X) \, \partial_{\tau} X^\nu \,=\, 0~.
\ee
This  is nothing but the familiar boundary condition, $\partial_{\sigma} \, X^{\mu} = 0 $, in flat space.
Similarly the Dirichlet boundary condition, $\partial_{\tau} X^{p}= \delta X^{p}  = 0$ should be replaced by $P^\tau_{p}= 0 $ in curved backgrounds.

 Thus the  generalized boundary conditions,  in the {\bf{open\, string\,  channel}}~\footnote{In the closed string channel, a pair of D-branes exchange one closed string. The boundaries of the worldsheet are at $\tau= \tau_{1}$ and $\tau= \tau_{2}$.  Closed-string channel is useful only when one analyses the boundary states.  To avoid possible confusion we will discuss physics  solely in the open-string picture
.}~ 
in a curved spacetime  are
\bea 
 {{\mathbf{Neumann:} }}  &&  ~~ P^{\mu}_\sigma  =  0 \\
 {{\mathbf{Dirichlet:} }}  && ~~ P^{p}_\tau = 0~.
\eea
We will see in the next section that these two boundary conditions are re-interpreted as two special linear combinations of currents at the boundary, namely:
\bea 
P^{\mu}_\sigma  &\leftrightarrow &\cJ_{Neumann} \\
P^{p}_\tau  &\leftrightarrow &\cJ_{Dirichlet}~.
\eea
As we shall see in Section~3 below when expressed in terms of the  free fields~\cite{Cheung:2003ym} the Neumann boundary for the free field indeed reduces to those in flat space.   At this point we would like to stress that the Neumann boundary conditions in curved space can only be realized by gluing the left-moving currents to the right-moving currents by an {{\bf{inner}}} automorphism of the group which inevitably leads to a gluing matrix with nontrivial field dependence.  This is to be contrast with the much studied D-branes arisen from field independent gluing conditions using group outer automorphisms of the group, notably, ${-\mathbb{I}}$.

 The  action of the Nappi-Witten model  is given by~\cite{Nappi:1993ie}:
\bea		\label{eq: sigma-model}
	\cS &=& \frac{1}{2\pi} \int \pd_+ u \, \pd_- v + \pd_+ v \, \pd_- u 
						+ \pd_+ a \,  \pd_- \ba + \pd_+ \ba \, \pd_- a   \nn
					&&\spa  + i\, H \, ( \ba \, \pd_- a \pd_+ u - a \pd_- \ba \, \pd_+ u )
\eea
where the background metric, $G$,  and the gauge invariant combination of Neveu-Schwarz B-field and gauge field,  $\cF = B+ F$,  are given by 
\bea \label{eq: G}
ds^2 &=&  du \, dv + da \, d\ba + 2\, i \, H\, ( \ba\,  da - a \, d\ba) du  \\
\label{eq: cF}
\cF &=&   2\, i \, H\, ( \ba\,  da - a\,  d\ba) \wedge  du
\eea
$H$ being  a  constant. 
Notice  however that we  are using a different gauge   from  that of~\cite{Nappi:1993ie}.  In going from the closed-string theory  to the open-string theory,   one has to specify a background gauge field,  $F$,   in addition to  the NS-background,   in order to fully specify an open string background.  It is the gauge invariant combination,  $\cF = B+ F$,  that  couples to the endpoints  of the open-strings.  Compare (eq. 2.6) in~\cite{Cheung:2003ym}, used here for the open strings,  and (eq. 3.4)  in~\cite{Cheung:2003ym},  for the closed strings: we had introduced a constant gauge field, $F = du\wedge dv +  da   \wedge d\ba$.    This necessity of using a different gauge from the original one adopted by Nappi and Witten in studying open string was  also noticed by the authors of~\cite{Dolan:2002px, Hashimoto:2004pb} when they studied other aspects of the Nappi-Witten model.   
We will also set $H=1$ in the later part of the paper.

We now vary the above action and study the boundary conditions at the endpoints of the string, $\sigma= 0$ and 
$\sigma = \pi$:
\bea
\label{eq: vBC}     \delta v ~ (-\pd_+ u  +\pd_-  u ) &=& 0 \\
\label{eq: uBC}  \delta u ~ (  -\pd_+ v  +\pd_- v + i\,H \, ( \ba \, \pd_- a - a \,\pd_- \ba  )  )
                                          &=& 0   \\
\label{eq: aBC}    \delta\bar{a}  ~(-\pd_+ \ba  +\pd_- \ba   -  i\, H\,  \ba \pd_+ u ) &=& 0 \\
\label{eq: abarBC}    \delta a ~ ( -\pd_+ a  +\pd_- a   +  i\, H \, a \pd_+ u ) &=& 0   
\eea 
We are looking for a solution that satisfies Neumann boundary condition in all four directions.  
So we demand  that  the four expressions inside the parenthesis vanish.

Using the definition of the canonical momentum, $P^\sigma$, one can show that the vanishing $P^\sigma$  is identical to requiring   all four expressions inside the parenthesis  in (\ref{eq: vBC}),  (\ref{eq: uBC}), (\ref{eq: aBC}), (\ref{eq: abarBC})  vanish.   This unambiguously established  the ``Neumannity'' of the above boundary conditions~\footnote{This is usually called ``mixed Neumann Dirichlet'' boundary condition in the literature, which we found misleading.}.


\section{Currents and their Gluing Conditions}

Nappi-Witten  model is also a WZW model in which case there are  infinite-dimensional  conserved currents on the closed string worldsheet.  In fact there are two independent sets of such currents, one for the left-movers and one for the right-movers, generating  the
 ${{\mathbf{G}(z)\times\mathbf{G}(\bz)}}$  isometry of the loop group.   The insertion of boundary implies that the left-moving and right-moving currents are no longer independent.
 They are related to each other at the boundary by a ``reflection matrix,''  $\cR$:
\be\label{eq: JcRJ}
J_L^a (z) +\cR^a _{\, b} ~  J_R^b (\bar{z})=0~.
\ee
$\cR^a _{\, b} $  are  the component of  a $z$ dependent linear map acting on  the Lie 
algebra.    
The gluing matrix, $\cR $,   encodes both Dirichlet and Neumann boundary conditions. In flat space $+1$ corresponds to a Neumann direction and $-1$ is Dirichlet.   Comparing with the expression of boundary condition~(\ref{eq: sigmaBC})  from the sigma model analysis and use $\cJ_{L} = -  \partial g g^{-1}$ and  
$\cJ_{L} =  g^{-1} \bar\partial g $   we obtain
\be \label{cR}
\cR^a _{\, b}  =\, \left[ (G+\cF)^{-1} (G-\cF ) \, C \right]^a _{\, b}
\ee
where $G_{ij}, B_{ij}$ denotes the components of the metric and the B-field
 in terms of left invariant form fields,
$ds^2 = G_{ij} \, (g\, dg^{-1})^i \, (g\, dg^{-1})^j$, $B=B_{ij} \, (g\, dg^{-1})^i \wedge (g\, dg^{-1})^j$.
 $C^a _{\, b}$ denotes the adjoint map  
 \be
T_a \, C^a _{\, b} \equiv Ad(g) \cdot T_b  =  g T_b g^{-1}~.
 \ee
This relation between the  background field,  $\cF$,  and the $\cR $ field is invertible as long as $C + \cR$ does not  have  a kernel,  in which  case we can write
\be  \label{eq: F-C}
\cF = G \,  (C - \cR) \, (C + \cR)^{-1}.
\ee 
In other words this  equation  serves to  relate the algebraic data with the geometric data of WZW models.    If a  metric, $G$, and a background field, $\cF$ are given then $\cR$ can be determined from it; and vice versa.    The insight gained from the sigma model analysis will enable us to find a consistent background for open strings easily.    This gives us considerable advantage over the purely algebraic manipulation as the   consistency of a particular  open-string background is not obvious in the closed-string picture.

Using this recipe we can directly compute the reflection matrix $\cR$ in Nappi-Witten model.    For  the  space-filling D3-brane\footnote{%
The currents are vectors on the tangent space  of the group manifold,
$\vec{\cJ} \equiv \cJ^+  \hat{J}^- + \cJ^-  \hat{J}^+ + \cT\,  \hat{J} + \cJ\,  \hat{T}$,  where 
$\hat{J}^-$,  $\hat{J}^+$, $\hat{J}$, $ \hat{T}$,  the generators of the algebra,  act as the basis for  the  tangent vectors  of  the group manifold.  To save typing we will drop the hats for the generators and  only use the calligraphy letters to denote chiral current components.}
it turns out to be~\cite{Stanciu:1998ak}
\be   	\label{eq: gluing}
\vec\cJ_L+ \cR  \vec\cJ_R = 0 ~~~  \rightarrow ~~~ \cR  =  Ad(e^{uJ})~.
\ee
where  $Ad(g)$  denotes  the adjoint action by the group element $g$,  $Ad(g)\, X = g X g^{-1}$.
The  boundary conditions  (\ref{eq: JcRJ})  are then explicitly  given by
\bea  \label{eq: JJ}
\cJ^\pm_L  + e^{\pm iu }\cJ^\pm_R &=& 0 \nn
\cJ_L + \cJ_R &=& 0 \nn
\cT_L + \cT_R &=& 0~.
\eea
The gluing matrices were   usually taken to be  outer automorphisms of the algebra, which give rise to a large body of  D-branes we  know and love.   Although the possibility of using inner automorphisms, and hence field-dependent gluing matrices,  has been noticed by many researchers~\cite{Alekseev:1998mc, Stanciu:1999id, Bachas:2000fr,  Lindstrom:2002mc,  Albertsson:2003va} and the constraint equation that $\cR$ has to satisfy in order to be an inner automorphism has been written down in~\cite{Albertsson:2003va},  in many  conformal field analysis,  due to  the lack of  geometric intuition,  the components of the gluing matrices  are quickly taken to    be  field-independent.  
Adjoint action by a group element  is of course an inner automorphism of the algebra.   The  element ${J}$ lives in the Cartan of the  Nappi-Witten algebra and it is the only element in the Cartan subalgebra that has nontrivial action on the rest of the algebra.   In a sense what we have proposed is the most natural generalization of the constant gluing matrix.  
										
Recall  that the Nappi-Witten algebra\footnote{%
This is a centrally-extended two-dimensional Euclidean algebra as it is manifest in the following 
way of writing it:
\be	
	[J, P_1 ] = P_2,  	\spa [ J, P_2 ] = -P_1, 	\spa [P_1 , P_2 ] = T, 	\spa [ T, * ] = 0 
\ee
In the note we have defined
 $J_+ = \frac{1}{\sqrt{2}} (P_1 - iP_2 )$ and  $J_- = \frac{1}{\sqrt{2}} (P_1 + iP_2 )$~.}
 is 
\be 	\label{eq: NWalgebra}
	[ J, J_+]  =  i J_+,         	\spa            [ J, J_- ]  =  -i J_-,          \spa    [J_+, J_- ] = iT~.
\ee
We shall parametrize a generic group element by
\be	\label{eq: groupelement}
 \displaystyle{g = e^ {aJ_+ + \ba J_-} e^{\, uJ + \, vT}}
 \ee
and use the definitions of $\cJ_+  =  - \pd_+ g g^{-1} $ and $\cJ_-  = g^{-1}  \pd_-  g$ 
to obtain the following sets of chiral currents:
\be   \label{eq: Jcomponents}
\begin{array}{rcl} 
 	-\cJ^+_L  & =&  \pd_+ \ba  + i \, \ba \pd_+ u \\
 	-\cJ^-_L  & =&  \pd_+ a  - i \, a \pd_+ u \\
 	-\cJ_L & =&  \pd_+ v + {i}\,  [ a \, \pd_+ \ba - \ba \, \pd_+ a  ] - a\, \ba \, \pd_+ u \\
 	-\cT_L & =&  \pd_+u
\end{array}
\spa
\begin{array}{rcl} 
	\cJ^+_R  & =&  e^{+iu} \pd_- \ba \\  
	\cJ^-_R  & =& e^{-iu} \pd_- a  \\
	\cJ_R    & =&   \pd_- v +  {i}[\, \ba \pd_- a - a \pd_- \ba\,] \\
	\cT_R   & =&   \pd_- u  
\end{array}
\ee
Let us remark that the minus sign in the definition of the left-moving current is not arbitrary: it is to ensure that  the left-moving and right-moving currents obey the same Kac-Moody algebra.

Now it is straightforward exercise to plug in the definitions of the left and right 
currents~(\ref{eq: Jcomponents}) above into the gluing conditions~(\ref{eq: gluing}) and obtain the following boundary conditions:
\bea 
\label{eq: Jv-Jv}			-\pd_+ u  + \pd_- u        &=& 0     \\
\label{eq: Ja-Ja}	-\pd_+ a  + \pd_- a +   i a\, \pd_+ u 	&=& 0    \\
\label{eq: Jba-Jba}      -\pd_+ \ba + \pd_-  \ba   -  i \ba\, \pd_+ u  &=& 0     \\
\label{eq: Ju-Ju}        -\pd_+ v + \pd_- v   + a\,  \ba\,   \pd_+ u
      +\frac{i}{2} \left[ \,\ba\, (\pd_+ +\pd_- )\, a  -  a \, (\pd_+ +\pd_- )\,\ba\right] &=&  0   
\eea
Compare these with the boundary conditions obtained earlier, (\ref{eq: vBC}),  (\ref{eq: uBC}),  (\ref{eq: aBC}),  (\ref{eq: abarBC}),  one sees that they indeed describe a D3-brane!  We have made use of (\ref{eq: aBC}) and (\ref{eq: abarBC}) to write (\ref{eq: uBC}) into the form of (\ref{eq: Ju-Ju}). 
So by now we have succeeded in showing that if we allow more general gluing conditions than those have been considered in the literature we can have a space-filling D-brane in  Nappi-Witten model.   The geometry of this D-brane is completely determined by the boundary conditions imposed on the open string endpoints.   In our case it is that of the plane-polarized gravitational waves supported by the  null Neveu-Schwarz three-from fluxes.  Whereas  the consistency of the open string background  is  apparent from the sigma-model analysis, the symmetry preserved by the underlying D-brane is most easily seen from the algebraic  study which  we shall turn to in the following section.  This concludes the classical analysis.  We shall turn our attention to the quantum algebra, the OPEs,  of the  currents at the boundary.  

						
\section{Current Algebra}

The gluing matrix relates the left-moving currents with the right-moving ones as in~(\ref{eq: JJ}).
This particular linear combination of the  chiral currents,  dictated by  the Neumann boundary condition,  will be shown  in this section,  to  obey  the same  Kac-Moody algebra as  the original  closed-string theory in the bulk.  This combination of chiral currents form a close algebra: an OPE of any  two  such ``boundary'' currents does not take one outside of  this set of  four  boundary currents.   The  other  linear combination with  a relative minus sign between the left and the  right  chiral currents, corresponding to Dirichlet boundary condition,  would not form a close algebra.    Alternatively in the closed string picture, we can 
construct  boundary operators,  $Q^a$,  as  the conserved charges of  these  boundary currents.   $Q^a$  will have to annihilate the boundary state~\cite{Cardy:1989ir} representing this D3-brane.  Any commutators $[Q^a, Q^b]$   will have to annihilate this very  boundary state.    This again  implies that it is consistent to impose Neumann  boundary condition on all of these four directions.   The underlying D3-brane, or the boundary state,  thus preserves half of  infinite dimensional ${{\mathbf{G}(z)\times\mathbf{G}(\bz)}}$   symmetries of the  WZW model.    This is analogous  to the BPS property  of  D-branes--preserving  half of the spacetime supersymmetry.  Given that the gluing condition is actually an inner automorphism of the algebra it is perhaps not surprising to see part of the original symmetry  get preserved after all.  
 
To prove our claims let us recall that the bulk currents satisfy the OPE:
\bea    \label{eq: OPEbetagamma}
          \cJ (z) \, \cJ^\pm (w) &\sim& \displaystyle \pm i \, \frac{ \cJ^\pm(z)}{z - w}   \nn
           \cJ^+(z) \, \cJ^-(w) &\sim& \displaystyle\frac{1}{(z-w)^2} +\, i \, \frac{\cT(z)}{z-w}  \\
           \cT(z) \, \cJ (w)  &\sim& \displaystyle\frac{ 1}{(z - w)^2}~. \nonumber
\eea
These  OPEs are easily realized  using  the   Wakimoto ``free-field'' representation introduced in~\cite{Cheung:2003ym}:  
%
\be
\bary{rclrcl}\label{eq: Jbetagamma}
-\cJ^+_L(z)      &=&  e^{-iu_{L}}\, \partial \gamma_L~~~~~~~~ &
\cJ^+_R(\bz)   &=&   e^{iu_{R}}\, \bar\beta_R  \\
-\cJ^-_L(z)       &=& e^{iu_{L}}\, \beta_L &
\cJ^-_R(\bz)  &=&  e^{-iu_{R}} \, \bpd \bar\gamma_R \\
-\cJ_L(z)         &=& \pd {v}_L &
\cJ_R(\bz)     &=&  \bpd {v}_R  \\
-\cT_L(z)        &=&  \pd u_L & 
\cT_R(\bz)   &=&  \bpd  u_R\\ \eary
\ee
where  $\beta = \pd \bar\gamma$ and $\bar\beta = \bpd \gamma$.
Using the following   free fields  contraction rules:
\bea \label{freefieldOPE}
u(z)       \,  {v}(w)              & \sim &  \ln (z-w)  \nn
\beta(z) \,  \gamma(w) & \sim & \frac{1}{z-w}~,
\eea
it is a straightforward exercise to verify that the chiral currents~(\ref{eq: Jbetagamma}) obey 
the OPEs given in~(\ref{eq: OPEbetagamma}).   

We can now proceed to verify that the linear combinations of the chiral currents in the left hand side of~(\ref{eq: JJ}) are consistent, {\em{i.e.}} the algebra closes onto itself.    
\bea \label{Jplus-Jminus}
&&[\cJ_L^+ (z) + e^{iu(z,\bz)} \cJ_R^+ (\bz) ][\cJ_L^- (w) + e^{-iu(w,\bw)} \cJ_R^- (\bw)]  \nn
&\sim&  \frac{1}{(z-w)^2} + \frac{1}{(\bz-\bw)^2}
 + \frac{i \cT_L(z)}{(z-w)}+ \frac{i \cT_R(\bz)}{(\bz-\bw)} 
\eea
\be
\label{J-plus}
[\cJ_L(z) + \cJ_R(\bz)] [\cJ_L^+ (w) + e^{iu(w,\bw)} \cJ_R^+(\bw)] 
\sim   i \left( \frac{\cJ^+_L(z)}{(z-w)} + \frac{ e^{iu(z,\bz)} \cJ^+_R(\bz)}{(\bz-\bw)}  
\right) 
\ee
\be
\label{J-minus}
[ \cJ_L(z) + \cJ_R(\bz)][\cJ_L^-(w) + e^{-iu(w,\bw)} \cJ_R^-(\bw)]  
\sim  -i \left(\frac{\cJ^-_L(z)}{(z-w)} + \frac{ e^{-iu(z,\bz)} \cJ^-_R(\bz)}{(\bz-\bw)}\right)  
\ee
\be
\label{T-J}
[ \cT_L(z)  +\cT_R(\bz) ] [\cJ_L(w) + \cJ_R(\bw)]
 \sim  \frac{1}{(z-w)^2}+  \frac{1}{(\bz-\bw)^2}
\ee
All other OPEs are regular.  
The right hand side of the equations are to be evaluated at the boundary, $z={\bar z}$.

Let us remark that the phase $e^{iu(z, \bz)}$ being non-chiral is essential for the closure of the above OPEs.
The contribution from 
$\cJ_L e^{iu(z, \bz)}$ is equal and opposite of that from  $\cJ_R e^{iu(z, \bz)}$:
\be
\cJ_L(z) e^{iu(w, \bw)} \sim \frac{-i  e^{iu(z, \bz)} } {z-w}  \spa \spa
\cJ_R(\bz)  e^{iu(w, \bw)}  \sim  \frac{i  e^{iu(z, \bz)} } {\bz-\bw}
\ee
which is the magic that enables the algebra to close.

 Notice however that the central charge of the system is the {\em{ total }}  of the left-moving and right-moving parts.  
One can also prove that the above currents are conserved and indeed generate the symmetry transformations advertised.  
This is left as an exercise to  the readers. 
 An interesting open question is to investigate the implication of this infinite dimensional symmetry present in the D3-branes on the BPS property of the object.  The interplay of these  extra infinite-dimensional symmetries  with the other (super)symmetries  in the target space  should also  be understood better and will be reported in a separate publication.

Furthermore, using the free fields we can rewrite the  boundary conditions~(\ref{eq: vBC}), (\ref{eq: uBC}), (\ref{eq: aBC}), (\ref{eq: abarBC}),   as 
\bea \label{freefieldBC}
\pd u_L -  \bpd  u_R  &=& 0 \\
\pd {v}_L  - \bpd {v}_R  &=& 0 \\
\pd \bar\gamma_L  -  e^{-iHu_R} \bpd (  e^{iHu_R} \bar\gamma_R)&=& 0 \\
e^{-iHu_L} \pd (e^{iHu_L} \gamma_L) - \bpd \gamma_R &=& 0 
\eea
where we have  re-instated the dependence on the field strength, $H$, a real parameter.
In the cases of the physical free  fields like  $u$ and ${v}$  they are the familiar Neumann boundary conditions.
The twisting (a phase factor)  in the other two equations is due to the fact that $\bar\gamma_L$ and $\gamma_R$ are not physical fields.    They are related to the physical ones,
$\bar\gamma_R$  and $\gamma_L$, respectively,  at the boundary by nontrivial phases,  $e^{iHu_R}$ and $e^{iHu_L} $,  respectively.    The readers are referred to Section~3.2 of~\cite{Cheung:2003ym} for a detailed discussion on how the two sets  of fields come about  in NW model.   This explains the form of the boundary conditions above.  

To conclude we have successfully shown that it is much more straightforward to study D-branes 
if one fully  exploits the dual  geometric and algebraic properties of WZW models. 
We demonstrate our technique by uncovering a  space-filling  D-branes in a particular WZW  model, that discovered by C.~Nappi and E.~Witten.    This D-brane  arises when  one  allows for  field-dependent gluing matrix.   The reflection matrix  is shown to be an adjoint action by a particular group element, $e^{uJ}$,  on the algebra.  These  gluing conditions on the  chiral currents is shown to be  identical to the Neumann  boundary condition of the open strings.  The merit of the method is that it allows the geometric property D-branes to remain  transparent by fully utilizing  the geometric aspect of WZW models.   In order for it to be useful however  the technique should be generalized.   On the conformal field theory side one would like to generalize the boundary states  to reflect  the  field-dependence in  gluing matrix.   
It is  also educational  to apply our techniques to the better understood models like  $SL(2)$ and $SU(2)$  to see if one gets new insight into these theories.  It also serves as a cross check for our techniques.


\section*{Acknowledgement}
We would like to thank Konstantin Savvidy for initial collaboration and many useful discussions throughout the project.  We would also like to thank Costas Bachas and Stephen Hwang for critical reading of the manuscript and useful discussions.  We are also indebted to Gang~Chen, Zhe-Yong~Fan, Congxin~Qiu  and Sibo~Zheng for carefully checking the algebras.
We have also benefited from discussions with  Louise Dolan and  Hiroshi Ooguri.  We are grateful to the referee for his/her careful reading of the manuscript and whose comment helps improving the presentation of the paper. 
E.~C.  would also like to thank Hai-Qing Lin of the Physics Department at  Chinese University of Hong Kong  for warm hospitality at the final  stage of the project.
E.~C. and L.~F. acknowledge support from the visitor programme at Perimeter Institute.
The research at Perimeter Institute is supported by NSERC.  
E.~C acknowledges support from the National Science Foundation of China under grant No.~0204131361   and Startup grants from Nanjing University as well as 985 Grant No.~020422420100 from the Chinese Government.


\addcontentsline{toc}{section}{Reference}
\begin{thebibliography} {99}

\bibitem{Dbranes}
  J.~Polchinski,
  ``Dirichlet-Branes and Ramond-Ramond Charges,''
  Phys.\ Rev.\ Lett.\  {\bf 75}, 4724 (1995)
  [arXiv:hep-th/9510017]\\
  P.~Horava,
  ``Strings On World Sheet Orbifolds,''
  Nucl.\ Phys.\ B {\bf 327}, 461 (1989).
  M.~B.~Green,
 ``Point - like states for type 2b superstrings,''
  Phys.\ Lett.\ B {\bf 329}, 435 (1994)
  [arXiv:hep-th/9403040]\\

\bibitem{Witten:WZW}
  E.~Witten,
  ``Nonabelian Bosonization In Two Dimensions,''
  Commun.\ Math.\ Phys.\  {\bf 92}, 455 (1984).
  
\bibitem{Nappi:1993ie}
C.~R.~Nappi and E.~Witten,
``A WZW model based on a nonsemisimple group,''
Phys.\ Rev.\ Lett.\  {\bf 71}, 3751 (1993)
[arXiv:hep-th/9310112]. 

\bibitem{Polchinski:1987tu}
  J.~Polchinski and Y.~Cai,
  ``Consistency Of Open Superstring Theories,''
  Nucl.\ Phys.\ B {\bf 296}, 91 (1988).

\bibitem{Abouelsaood:1986gd}
  A.~Abouelsaood, C.~G.~.~Callan, C.~R.~Nappi and S.~A.~Yost,
   ``Open Strings In Background Gauge Fields,''
  Nucl.\ Phys.\ B {\bf 280}, 599 (1987).
  
\bibitem{Ooguri:1996ck}
  H.~Ooguri, Y.~Oz and Z.~Yin,
  ``D-branes on Calabi-Yau spaces and their mirrors,''
  Nucl.\ Phys.\ B {\bf 477}, 407 (1996)
  [arXiv:hep-th/9606112].
  
\bibitem{Fuchs:1997fu}
  J.~Fuchs and C.~Schweigert,
  ``Branes: From free fields to general backgrounds,''
  Nucl.\ Phys.\ B {\bf 530}, 99 (1998)
  [arXiv:hep-th/9712257].

  
\bibitem{Maldacena:2001ky}
  J.~M.~Maldacena, G.~W.~Moore and N.~Seiberg,
  ``Geometrical interpretation of D-branes in gauged WZW models,''
  JHEP {\bf 0107}, 046 (2001)
  [arXiv:hep-th/0105038].
  
\bibitem{Giveon:2001uq}
A.~Giveon, D.~Kutasov and A.~Schwimmer,
``Comments on D-branes in AdS(3),''
Nucl.\ Phys.\ B {\bf 615}, 133 (2001)
[arXiv:hep-th/0106005].

\bibitem{Lee:2001gh}
P.~Lee, H.~Ooguri and J.~w.~Park,
``Boundary states for AdS(2) branes in AdS(3),''
Nucl.\ Phys.\ B {\bf 632}, 283 (2002)
[arXiv:hep-th/0112188].

\bibitem{Ponsot}
B.~Ponsot, V.~Schomerus and J.~Teschner,
``Branes in the Euclidean AdS(3),''
JHEP {\bf 0202}, 016 (2002)
[arXiv:hep-th/0112198].

\bibitem{Klimcik:1996hp}
  C.~Klimcik and P.~Severa,
  ``Open strings and D-branes in WZNW models,''
  Nucl.\ Phys.\ B {\bf 488}, 653 (1997)
  [arXiv:hep-th/9609112].
   
\bibitem{Stanciu:1998ak}
S.~Stanciu and A.~A.~Tseytlin,
``D-branes in curved spacetime: Nappi-Witten background,''
JHEP {\bf 9806}, 010 (1998)
[arXiv:hep-th/9805006].

\bibitem{Alekseev:1998mc}
A.~Y.~Alekseev and V.~Schomerus,
``D-branes in the WZW model,''
Phys.\ Rev.\ D {\bf 60}, 061901(R) (1999)
[arXiv:hep-th/9812193].

\bibitem{Stanciu:1999nx}
  S.~Stanciu,
  ``D-branes in an AdS(3) background,''
  JHEP {\bf 9909}, 028 (1999)
  [arXiv:hep-th/9901122].
  
\bibitem{Alekseev:1999bs}
  A.~Y.~Alekseev, A.~Recknagel and V.~Schomerus,
  ``Non-commutative world-volume geometries: Branes on SU(2) and fuzzy   spheres,''
  JHEP {\bf 9909}, 023 (1999)
  [arXiv:hep-th/9908040].
  
\bibitem{Felder:1999ka}
G.~Felder, J.~Frohlich, J.~Fuchs and C.~Schweigert,
``The geometry of WZW branes,''
J.\ Geom.\ Phys.\  {\bf 34}, 162 (2000)
[arXiv:hep-th/9909030].

\bibitem{Stanciu:1999id}
  S.~Stanciu,
 ``D-branes in group manifolds,''
  JHEP {\bf 0001}, 025 (2000)
  [arXiv:hep-th/9909163].
  
\bibitem{Figueroa-O'Farrill:1999ie}
  J.~M.~Figueroa-O'Farrill and S.~Stanciu,
 ``More D-branes in the Nappi-Witten background,''
  JHEP {\bf 0001}, 024 (2000)
  [arXiv:hep-th/9909164].

\bibitem{Malda-Oog}
J.~M.~Maldacena and H.~Ooguri,
``Strings in AdS(3) and SL(2,R) WZW model. I,''
J.\ Math.\ Phys.\  {\bf 42}, 2929 (2001)
[arXiv:hep-th/0001053].


\bibitem{Pawelczyk:2000ah}
  J.~Pawelczyk,
  ``SU(2) WZW D-branes and their noncommutative geometry from DBI action,''
  JHEP {\bf 0008}, 006 (2000)
  [arXiv:hep-th/0003057].
 

\bibitem{Ishikawa:2000ez}
  H.~Ishikawa and S.~Watamura,
  ``Free field realization of D-brane in group manifold,''
  JHEP {\bf 0008}, 044 (2000)
  [arXiv:hep-th/0007141].
  

\bibitem{Bachas:2000fr}
C.~Bachas and M.~Petropoulos,
``Anti-de-Sitter D-branes,''
JHEP {\bf 0102}, 025 (2001)
[arXiv:hep-th/0012234].

\bibitem{Stanciu:2001vw}
  S.~Stanciu,
  ``An illustrated guide to D-branes in SU(3),''
  arXiv:hep-th/0111221.

\bibitem{Stanciu:2001ug}
  S.~Stanciu,
  ``A geometric approach to D-branes in group manifolds,''
  Fortsch.\ Phys.\  {\bf 50} (2002) 980
  [arXiv:hep-th/0112130].

\bibitem{Sarkissian:2002ie}
  G.~Sarkissian,
  ``Non-maximally symmetric D-branes on group manifold in the Lagrangian approach,''
  JHEP {\bf 0207}, 033 (2002)
  [arXiv:hep-th/0205097].
  
\bibitem{Lindstrom:2002mc}
  U.~Lindstrom, M.~Rocek and P.~van Nieuwenhuizen,
  ``Consistent boundary conditions for open strings,''
  Nucl.\ Phys.\ B {\bf 662}, 147 (2003)
  [arXiv:hep-th/0211266].
  
\bibitem{Lindstrom:2002vp}
  U.~Lindstrom and M.~Zabzine,
  ``D-branes in N = 2 WZW models,''
  Phys.\ Lett.\ B {\bf 560}, 108 (2003)
  [arXiv:hep-th/0212042].
    
\bibitem{Stanciu:2003bn}
S.~Stanciu and J.~Figueroa-O'Farrill,
``Penrose limits of Lie branes and a Nappi-Witten braneworld,''
JHEP {\bf 0306}, 025 (2003)
[arXiv:hep-th/0303212].

\bibitem{Albertsson:2003va}
  C.~Albertsson, U.~Lindstrom and M.~Zabzine,
  ``Superconformal boundary conditions for the WZW model,''
  JHEP {\bf 0305}, 050 (2003)
  [arXiv:hep-th/0304013].
  
\bibitem{Sarkissian:2003yw}
  G.~Sarkissian and M.~Zamaklar,
  ``Symmetry breaking, permutation D-branes on group manifolds: Boundary states and geometric description,''
  Nucl.\ Phys.\ B {\bf 696}, 66 (2004)
  [arXiv:hep-th/0312215].
  
\bibitem{Giribet:2007wp}
  G.~Giribet, A.~Pakman and L.~Rastelli,
  ``Spectral Flow in AdS(3)/CFT(2),''
  JHEP {\bf 0806}, 013 (2008)
  [arXiv:0712.3046 [hep-th]].

\bibitem{Baron:2008qf}
  W.~H.~Baron and C.~A.~Nunez,
  ``Fusion rules and four-point functions in the SL(2,R) WZNW model,''
  Phys.\ Rev.\  D {\bf 79}, 086004 (2009)
  [arXiv:0810.2768 [hep-th]].

   

\bibitem{Blau:2001ne}
M.~Blau, J.~Figueroa-O'Farrill, C.~Hull and G.~Papadopoulos,
``A new maximally supersymmetric background of IIB superstring theory,''
JHEP {\bf 0201}, 047 (2002)
[arXiv:hep-th/0110242].
\bibitem{Metsaev:2001bj}
R.~R.~Metsaev,
 ``Type IIB Green-Schwarz superstring in plane wave Ramond-Ramond background,''
Nucl.\ Phys.\ B {\bf 625}, 70 (2002)
[arXiv:hep-th/0112044].
\bibitem{Berenstein:2002jq}
D.~Berenstein, J.~M.~Maldacena and H.~Nastase,
``Strings in flat space and pp waves from N = 4 super Yang Mills,''
JHEP {\bf 0204}, 013 (2002)
[arXiv:hep-th/0202021].

\bibitem{Metsaev:2002re}
R.~R.~Metsaev and A.~A.~Tseytlin,
 ``Exactly solvable model of superstring in plane wave Ramond-Ramond background,''
Phys.\ Rev.\ D {\bf 65}, 126004 (2002)
[arXiv:hep-th/0202109].


\bibitem{ppbranes}
M.~Billo and I.~Pesando,
``Boundary states for GS superstrings in an Hpp wave background,''
Phys.\ Lett.\ B {\bf 536}, 121 (2002)
[arXiv:hep-th/0203028].\\
%
C.~S.~Chu and P.~M.~Ho,
``Noncommutative D-brane and open string in pp-wave background with B-field,''
Nucl.\ Phys.\ B {\bf 636}, 141 (2002)
[arXiv:hep-th/0203186].\\
%
A.~Dabholkar and S.~Parvizi,
``Dp branes in pp-wave background,''
Nucl.\ Phys.\ B {\bf 641}, 223 (2002)
[arXiv:hep-th/0203231].\\
%
  D.~Berenstein, E.~Gava, J.~M.~Maldacena, K.~S.~Narain and H.~Nastase,
  ``Open strings on plane waves and their Yang-Mills duals,''
  arXiv:hep-th/0203249. \\
%
A.~Kumar, R.~R.~Nayak and S.~Siwach,
``D-brane solutions in pp-wave background,''
Phys.\ Lett.\ B {\bf 541}, 183 (2002)
[arXiv:hep-th/0204025].\\
%
K.~Skenderis and M.~Taylor,
``Branes in AdS and pp-wave spacetimes,''
JHEP {\bf 0206}, 025 (2002)
[arXiv:hep-th/0204054].\\
%
H.~Takayanagi and T.~Takayanagi,
 ``Open strings in exactly solvable model of curved space-time and  pp-wave limit,''
JHEP {\bf 0205}, 012 (2002)
[arXiv:hep-th/0204234].\\
%
P.~Bain, P.~Meessen and M.~Zamaklar,
``Supergravity solutions for D-branes in Hpp-wave backgrounds,''
Class.\ Quant.\ Grav.\  {\bf 20}, 913 (2003)
[arXiv:hep-th/0205106].\\
%
M.~Alishahiha and A.~Kumar,
``D-brane solutions from new isometries of pp-waves,''
Phys.\ Lett.\ B {\bf 542}, 130 (2002)
[arXiv:hep-th/0205134].\\
%
O.~Bergman, M.~R.~Gaberdiel and M.~B.~Green,
``D-brane interactions in type IIB plane-wave background,''
JHEP {\bf 0303}, 002 (2003)
[arXiv:hep-th/0205183].\\
%
Y.~Michishita,
``D-branes in NSNS and RR pp-wave backgrounds and S-duality,''
JHEP {\bf 0210}, 048 (2002)
[arXiv:hep-th/0206131].\\
%
A.~Biswas, A.~Kumar and K.~L.~Panigrahi,
``p--p' branes in pp-wave background,''
Phys.\ Rev.\ D {\bf 66}, 126002 (2002)
[arXiv:hep-th/0208042].\\
%
O.~J.~Ganor and U.~Varadarajan,
``Nonlocal effects on D-branes in plane-wave backgrounds,''
JHEP {\bf 0211}, 051 (2002)
[arXiv:hep-th/0210035].\\
%
J.~F.~Morales,
``String theory on Dp-plane waves,''
JHEP {\bf 0301}, 008 (2003)
[arXiv:hep-th/0210229].\\
%
R.~R.~Nayak,
``D-branes at angle in pp-wave background,''
Phys.\ Rev.\ D {\bf 67}, 086006 (2003)
[arXiv:hep-th/0210230].\\
Y.~Hikida and S.~Yamaguchi,
``D-branes in pp-waves and massive theories on worldsheet with boundary,''
JHEP {\bf 0301}, 072 (2003)
[arXiv:hep-th/0210262].\\
%
K.~Skenderis and M.~Taylor,
``Open strings in the plane wave background. I: Quantization and symmetries,''
Nucl.\ Phys.\ B {\bf 665}, 3 (2003)
[arXiv:hep-th/0211011].\\
%
M.~R.~Gaberdiel and M.~B.~Green,
 ``The D-instanton and other supersymmetric D-branes in IIB plane-wave string theory,''
Annals Phys.\  {\bf 307}, 147 (2003)
[arXiv:hep-th/0211122]. \\
%
K.~Skenderis and M.~Taylor,
``Open strings in the plane wave background. II: Superalgebras and spectra,''
JHEP {\bf 0307}, 006 (2003)
[arXiv:hep-th/0212184].\\
%
S.~j.~Hyun, J.~Park and H.~j.~Shin,
``Covariant description of D-branes in IIA plane-wave background,''
Phys.\ Lett.\ B {\bf 559}, 80 (2003)
[arXiv:hep-th/0212343].\\
%
Y.~Hikida, H.~Takayanagi and T.~Takayanagi,
``Boundary states for D-branes with traveling waves,''
JHEP {\bf 0304}, 032 (2003)
[arXiv:hep-th/0303214].\\
%
C.~V.~Johnson and H.~G.~Svendsen,
``D-brane anti-D-brane forces in plane wave backgrounds: A fall from  grace,''
JHEP {\bf 0305}, 055 (2003)
[arXiv:hep-th/0303255].\\
%
M.~R.~Gaberdiel, M.~B.~Green, S.~Schafer-Nameki and A.~Sinha,
``Oblique and curved D-branes in IIB plane-wave string theory,''
JHEP {\bf 0310}, 052 (2003)
[arXiv:hep-th/0306056].\\
%
Y.~Kim and J.~Park,
``Boundary states in IIA plane-wave background,''
Phys.\ Lett.\ B {\bf 572}, 81 (2003)
[arXiv:hep-th/0306282]. \\
%
C.~Y.~Park,
``Open string spectrum in pp-wave background,''
J.\ Korean Phys.\ Soc.\  {\bf 44}, 235 (2004)
[arXiv:hep-th/0308151]. \\
%
G.~Sarkissian and M.~Zamaklar,
``Diagonal D-branes in product spaces and their Penrose limits,''
JHEP {\bf 0403}, 005 (2004)
[arXiv:hep-th/0308174]. \\
%
C.~P.~Bachas and M.~R.~Gaberdiel,
``World-sheet duality for D-branes with travelling waves,''
JHEP {\bf 0403}, 015 (2004)
[arXiv:hep-th/0310017].\\
%
K.~S.~Cha, B.~H.~Lee and H.~S.~Yang,
``A complete classification of D-branes in type IIB plane wave  background,''
JHEP {\bf 0403}, 058 (2004)
[arXiv:hep-th/0310177].\\
%
M.~Sakaguchi and K.~Yoshida,
``D-branes of covariant AdS superstrings,''
Nucl.\ Phys.\ B {\bf 684}, 100 (2004)
[arXiv:hep-th/0310228].\\
%
S.~F.~Hassan, R.~R.~Nayak and K.~L.~Panigrahi,
``D-branes in the NS5 near-horizon pp-wave background,''
[arXiv:hep-th/0312224].\\
%
Y.~Hikida,
``Boundary states in the Nappi-Witten model,''
arXiv:hep-th/0409185.\\
%
G.~D'Appollonio and E.~Kiritsis,
``D-branes and BCFT in Hpp-wave backgrounds,''
arXiv:hep-th/0410269.\\
%
P.~Mukhopadhyay,
``Non-BPS D-branes in light-cone Green-Schwarz formalism,''
[arXiv:hep-th/0411103]. 

\bibitem{Cheung:2003ym}
Y.~K.~E.~Cheung, L.~Freidel and K.~Savvidy,
``Strings in gravimagnetic fields,''
JHEP {\bf 0402}, 054 (2004)
[arXiv:hep-th/0309005].

\bibitem{D'Appollonio:2003dr}
G.~D'Appollonio and E.~Kiritsis,
``String interactions in gravitational wave backgrounds,''
Nucl.\ Phys.\ B {\bf 674}, 80 (2003)
[arXiv:hep-th/0305081].

\bibitem{Dolan:2002px}
  L.~Dolan and C.~R.~Nappi,
   ``Noncommutativity in a time-dependent background,''
  Phys.\ Lett.\ B {\bf 551}, 369 (2003)
  [arXiv:hep-th/0210030].

\bibitem{Hashimoto:2004pb}
  A.~Hashimoto and K.~Thomas,
  ``Dualities, twists, and gauge theories with non-constant non-commutativity,''
  JHEP {\bf 0501}, 033 (2005)
  [arXiv:hep-th/0410123].

\bibitem{Cardy:1989ir}
  J.~L.~Cardy,
  ``Boundary Conditions, Fusion Rules And The Verlinde Formula,''
  Nucl.\ Phys.\ B {\bf 324}, 581 (1989).

\end {thebibliography}
\end {document}